\documentclass{egpubl}
\usepackage{pg2024}

\JournalSubmission    
\CGFccby

\usepackage[T1]{fontenc}
\usepackage{dfadobe}  

\usepackage{amsmath}
\usepackage{amssymb}
\usepackage{acronym}
\acrodef{MLP}[MLP]{Multi-layer Perceptron}
\acrodef{GS}[GS]{Gaussian Splatting}
\usepackage[subtle]{savetrees}

\usepackage{mathrsfs}

\usepackage{xcolor}

\usepackage{lipsum}
\setlipsum{%
  par-before = \begingroup\color{lightgray},
  par-after = \endgroup
}

\usepackage{xcolor}
\newcommand{\rev}[1]{{\textcolor{black}{#1}}}

\usepackage{cite}  
\BibtexOrBiblatex
\electronicVersion
\PrintedOrElectronic
\ifpdf \usepackage[pdftex]{graphicx} \pdfcompresslevel=9
\else \usepackage[dvips]{graphicx} \fi

\usepackage{egweblnk} 

\title[$\mathscr{G}$-Style: Stylized Gaussian Splatting]%
      {$\mathscr{G}$-Style: Stylized Gaussian Splatting}

\author[Kovács et al.]
{\parbox{\textwidth}{\centering Áron Samuel Kovács \orcid{0000-0002-0849-9032}, Pedro Hermosilla, and Renata G.~Raidou \orcid{0000-0003-2468-0664}
        }
         \\
  {\parbox{\textwidth}{\centering TU Wien, Austria}
         }
 }

\begin{document}

\teaser{
\includegraphics[width=1\linewidth]{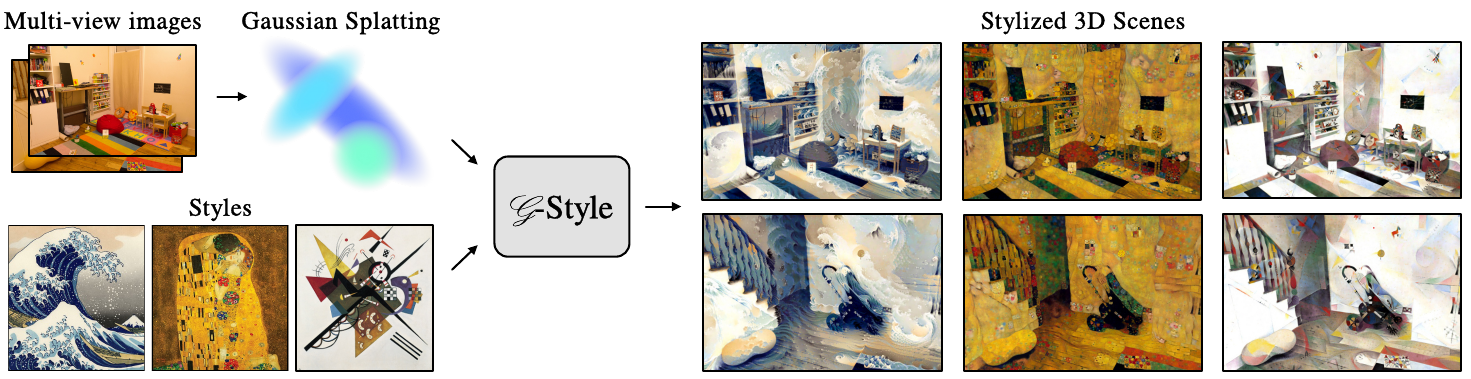}
\centering
 \caption{\textbf{$\mathscr{G}$-Style:} \rev{Our method takes a 3D scene, represented using Gaussian Splatting, and a style image exemplar as input, and generates a stylized version of the scene that closely matches the visual style of the exemplar. By modifying the geometry of the scene and designing losses that capture style patterns at different scales, we achieve high-quality stylized scenes efficiently, with results generated in just a few minutes.}}
\label{fig:teaser}
}

\maketitle
\begin{abstract}
\rev{We introduce $\mathscr{G}$-Style, a novel algorithm designed to transfer the style of an image onto a 3D scene represented using Gaussian Splatting.}
Gaussian Splatting is a powerful 3D representation for novel view synthesis, as---compared to other approaches based on Neural Radiance Fields---it provides fast scene renderings and user control over the scene.
Recent pre-prints have demonstrated that the style of Gaussian Splatting scenes can be modified using an image exemplar. 
However, since the scene geometry remains fixed during the stylization process, \rev{current solutions fall short of producing satisfactory results.}
Our algorithm aims to address these limitations by following a three-step process: In a pre-processing step, we remove undesirable Gaussians with large projection areas or highly elongated shapes.
Subsequently, we combine several losses carefully designed to preserve different scales of the style in the image, while \rev{maintaining as much as possible the integrity of the original scene content.}
During the stylization process and following the original design of Gaussian Splatting, we split Gaussians where additional detail is necessary within our scene by tracking the gradient of the stylized color.
Our experiments demonstrate that $\mathscr{G}$-Style generates high-quality stylizations \rev{within just a few minutes, outperforming} existing methods both qualitatively and quantitatively.

\ccsdesc[500]{Computing methodologies~Computer graphics}
\ccsdesc[500]{Computing methodologies~Artificial intelligence}
\ccsdesc[500]{Computing methodologies~Neural networks}

\printccsdesc   
\end{abstract}  

\section{Introduction}

While humans excel at creating paintings with specific contents and styles, \rev{this task has proven challenging for computers to replicate.}
With the advent of neural networks---and in particular, Convolutional Neural Networks---\rev{algorithms have been developed to transfer the style of one image onto another}~\cite{gatys2015styletransfer}.
\rev{In this process, a \textit{content image} refers to the original image whose subject matter we aim to retain, while a style image is the one whose artistic style we want to apply to the content image.
These algorithms enabled the modification of the style of an image while preserving its content by matching the statistical properties of the embeddings of both content and style images, as obtained from a deep neural network.}

With the appearance of novel view synthesis methods based on neural networks (NeRFs)~\cite{mildenhall2020nerf}, researchers turned their attention to applying style transfer techniques to entire 3D scenes.
Style transfer for 3D scenes aims to generate novel views of a scene from a finite number of images of the same scene with a particular style specified by a style image examplar.
To succeed in this task, the style transfer method should ensure multi-view consistency between views to provide a smooth navigation experience.
Despite the high-quality results provided by 2D style transfer methods, the same algorithms were not able to provide \textit{consistent styles} across 3D scene views~\cite{Huang_2021_ICCV, Huang_2022_CVPR, Mu_2022_CVPR, nguyen2022snerf}.
Therefore, they have been deemed unfit for 3D style transfer.
The limitations of these works have been addressed by several methods~\cite{Huang_2022_CVPR,nguyen2022snerf,zhang2022arf}---either by modifying the colors of the pre-trained NeRF representations or by techniques such as color transfer~\cite{zhang2022arf}, provided that the original methods were able to support view-dependent effects.
Still, NeRF-based approaches require \textit{large training} and \textit{rendering times}.

Recently, \ac{GS} has \rev{established itself as an active area of research~\cite{chen2024gssurvey, fei2024gssurvey, wu2024gssurvey} by demonstrating} that it can excel in view synthesis quality while requiring significantly less time for optimization and rendering.
Two recent pre-prints~\cite{liu2023stylegaussian,saroha2024gaussian} have suggested using \ac{GS} as the main scene representation and modifying the color of these to obtain stylized renderings of the scene.
These methods do not alter the position and shape of the Gaussian representing the scene, which results in a \textit{low resolution} in some areas of the scene and, hence, \textit{low-resolution stylized colors}.
In addition to the aforementioned problems, all 3D style transfer methods only focus on transferring \textit{high-frequency patterns} from the style image, such as brush strokes or color statistics.
However, the notion of the style of an image is \rev{more nuanced and can significantly} differ from image to image.
In some cases, like in Figure~\ref{fig:teaser}, the style is mostly defined by large patterns that existing methods might miss---thus, reducing \rev{their effectiveness in transferring the intended style.}

To overcome all these limitations, we introduce $\mathscr{G}$-Style, a novel method to transfer the style of an image onto a 3D scene represented using Gaussian Splatting. 
Our approach requires only a few minutes to optimize and delivers a high-quality stylized \ac{GS} representation of the scene.
Our method comprises several steps: 
First, we pre-process the scene represented with a set of Gaussians to ensure a uniform coverage. 
Then, our method starts the stylization process by updating the color associated with each Gaussian. 
\rev{During the stylization process, we enhance the resolution by splitting Gaussians with high color gradients, adding finer details where necessary.}
Due to the sparse nature of the scene and by only representing diffuse surfaces, our method provides view consistency by construction.
Moreover, our dual loss function enables the algorithm to capture both high-frequency and low-frequency patterns in the style image, therefore generating high-quality 3D scene renderings that \rev{reproduce a large variety of artistic styles.}
Our extensive evaluation demonstrates that our approach outperforms existing methods in style transfer quality and rendering time.
\section{Related Work}


\noindent \textbf{2D Style Transfer.}
Gatys et al.~\cite{gatys2015styletransfer} first introduced a method for neural style transfer, where the artistic style of a provided image is used to visually reconstruct the content of another image.
The method is based on iterative optimization to match the output and second-order statistics, expressed as the Gram matrices of hidden layers of a pre-trained network.
If preserving the content is not necessary, the method can synthesize a texture resembling the provided style image, initializing the process with a noise image~\cite{gatys2015texturesynthesis} and disregarding the part of the original loss function that is responsible for content preservation.

This approach has been since refined to enable transferring features on multiple scales~\cite{zhao2022stsgan}, by utilizing Generative Adversarial Networks (GANs) in a course-to-fine fashion~\cite{jetchev2017texturesynthesisgan} or diffusion models~\cite{zhang2023diffusionstyletransfer, wang2023stylediffusion, chung2024style}. 
Also, different ways of capturing and matching the style statistics have been proposed~\cite{liao2017imageanalogylossnn, gu2018arbitrarystyleloss, kolkin2019styleoptimaltransportlossnn}.
All aforementioned methods aim to improve style details by introducing pen or brush strokes, which were too blurry in the original work of Gatys et al., or by better preserving semantic consistency.

Instead of relying on matching extracted style statistics, searching for nearest neighbors in the feature space and minimizing distances between them is another option for transferring style~\cite{kolkin2019styleoptimaltransportlossnn, chen2016styletransferfastnn, liao2017imageanalogylossnn, li2016combiningmarkovnn, zhang2022arf}. 
This strategy can lead to significant improvements in transferring high-frequency features, by avoiding the averaging of features that match possibly multi-modal style statistics.

Existing methods can be further subdivided based on whether they iteratively modify an image, just like in the original work of Gatys et al., or explicitly minimize an objective function with a single feed-forward pass~\cite{chen2016styletransferfastnn, huang2017arbitrarytransferff, an2021artflowff}.
While the feed-forward-based approaches are orders of magnitude faster, their results are generally of lower quality than the slower optimization-based techniques.

\noindent \textbf{3D Style Transfer.}
3D style transfer refers to modifying the appearance of a 3D object or a scene so that when viewed from different angles, it matches the style of a given exemplar.
To reuse a 2D style transfer method, one needs to extract and employ a 2D image of a given scene. 
This can be either obtained by utilizing a differentiable renderer~\cite{mordvintsev2018styletransferrender, zhang2022arf}, 
by slicing the 3D volume~\cite{henzler2020neuraltexture, gutierrez2019volumetexture, zhao2022stsgan, chen2010high, kopf2007solid}, 
or by directly working on the surface manifold of textured meshes~\cite{kovacs2024surface}.

Existing approaches usually rely on a single way of representing objects and whole scenes, which is also the main point of influence for their performance.
The approach of Cao et al.~\cite{cao2020styletransferpointclouds} uses point clouds, yielding holes due to measurement errors during scanning and being unsuitable for representing surfaces and real-world scenes.
Conversely, Mordvintsev et al.~\cite{mordvintsev2018styletransferrender} and Hoellein et al.~\cite{hollein2022styletransferrender} use textured meshes, which can be reconstructed from the aforementioned point clouds.
Due to the discrete and restrictive nature of mesh representations, these approaches can suffer from different types of artifacts.

Lately, Neural Radiance Fields (NeRFs)~\cite{mildenhall2020nerf} and Gaussian Splatting~\cite{kerbl3Dgaussians} have become very prominent in reconstructing objects and scenes.
Unlike point clouds and meshes, NeRFs and Gaussians are soft volumetric representations.
Naturally, these representations can be utilized for scene reconstruction or style transfer.
Like in the 2D case, these style transfer methods can be subdivided into two categories: zero-shot and iterative methods.
Zero-shot methods~\cite{liu2023instant, liu2023stylerf, xu2024styledyrf} focus on matching colors, relighting, or transferring details on a small scale to achieve multi-view consistency.
This limitation arises from their inability to quickly and consistently embed large features into scenes. 
Thus, these methods struggle to synthesize large patterns that span significant portions of a given scene.

On the other hand, iterative methods leverage rendering scenes from multiple viewpoints.
Hence, they can construct large-scale patterns that remain consistent across different views, while any inconsistencies are corrected during the training process.
Zhang et al.~\cite{zhang2022arf} propose a new loss based on nearest neighbor feature matching (NNFM) which better preserves details and also optimizes backpropagation by deferring the changes to the network after first computing the losses for full-resolution images. 
Zhang et al.~\cite{zhang2024coarf} further improve this work by enabling the stylization of specific objects and introducing a semantic-aware version of NNFM.
These methods suffer from limited user control and require significant time to optimize---limitations that can be addressed by a novel view synthesis alternative: Gaussian Splatting~\cite{kerbl3Dgaussians}.

In this work, we focus on style transfer in scenes represented with the help of 3D Gaussian Splatting, allowing us to obtain a stylized scene representation in a few minutes and supporting real-time inspection of the result.
Two concurrent methods address the same problem as we do from different angles. 
The first, StyleGaussian, embeds 2D VGG scene features into reconstructed 3D Gaussians, transforms them according to a style, and decodes them onto a stylized image~\cite{liu2023stylegaussian}.
The second, GSS, employs pre-trained Gaussians conditioned on a style image to obtain stylized views of complex 3D scenes with spatial consistency~\cite{saroha2024gaussian}.
However, these methods do not change the geometry of the scene during optimization resulting in low-resolution stylized colors.
As we demonstrate later, our method achieves higher-quality scene representations that preserve large patterns from the style image.
\begin{figure*}
  \includegraphics[width=\linewidth]{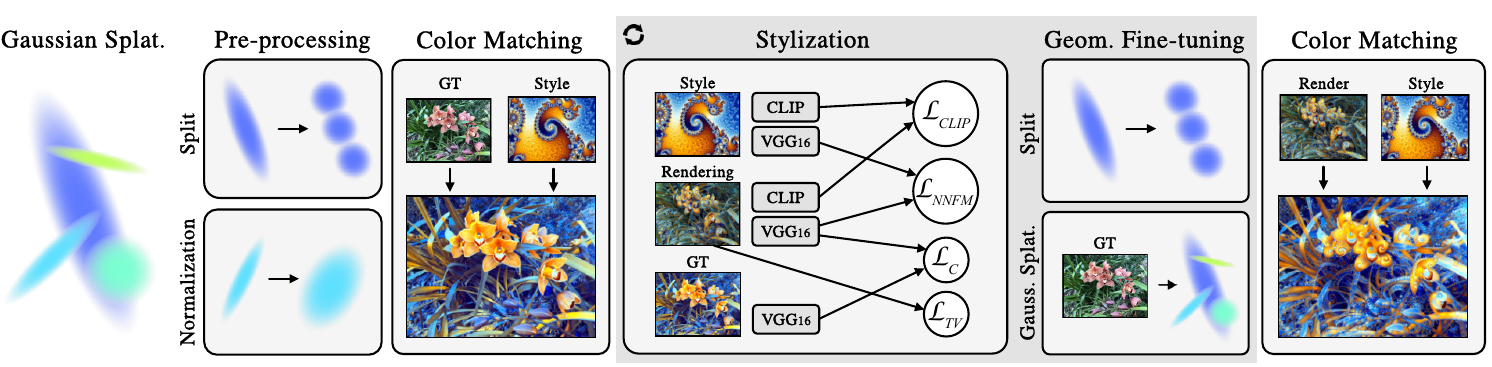}
  \caption{\textbf{Overview of our method:} We take a 3D scene represented using Gaussian Splatting and pre-process it to subdivide large Gaussians and normalize elongated ones. 
  Initially, we perform a color matching between the ground truth images and the style image.  
  Subsequently, we start the iterative stylization process.
  First, we optimize the colors of the Gaussians using multiple losses to capture style patterns at different scales while preserving the content of the scene. 
  Then, we fine-tune the geometry of the scene and add details for Gaussians with a large gradient of the stylized color. 
  We repeat the stylization and geometry fine-tuning steps until convergence.  
  At the end, we perform an additional color matching step between the renderings of the resulting scene and the style image.}
  \label{fig:overview}
\end{figure*}

\section{Background: NeRFs and Gaussian Splatting}

In this section, we briefly describe the two most commonly used 3D representations for novel view synthesis, NeRFs~\cite{mildenhall2020nerf} and Gaussian Splatting~\cite{kerbl3Dgaussians}.

\noindent \textbf{NeRFs.} 
Neural radiance fields have revolutionized the field of novel view synthesis by introducing a new scene representation, and an optimization algorithm to train this representation only from images.
The outgoing radiance at any point $x$ in the scene for any view direction $v$ is modeled by a parametric model $\phi_{\theta}(x, v)$ with parameters $\theta$.
This model is usually a \ac{MLP}, which is chosen due to the universality provided by this type of model.
To train this model, the scene is rendered from multiple views using the volume rendering algorithm~\cite{max1995optical}.
By comparing the generated image $\mathcal{I_{r}}$ to a real picture of the scene $\mathcal{I}_{gt}$, gradients can be computed for the parameters $\theta$ through the rendering operation and the scene representation can be updated to match the ground truth images.
Despite the high-quality images generated by NeRFs, the control provided to the user is limited, and they demand extensive rendering times due to the multiple evaluations required to compute the value of a pixel.

\noindent \textbf{Gaussian Splatting.} 
Gaussian Splatting~\cite{kerbl3Dgaussians} has emerged as a viable alternative to address the main limitations of NeRFs: limited scene control and low rendering speed.
To reconstruct a real-life scene from ground truth images $\mathcal{I}_{gt}$, Kerbl et al. use the Structure from Motion algorithm~\cite{ullman1979interpretation} to obtain camera poses for each image and a spare set of initial 3D points.
The 3D points are then transformed into Gaussian functions, each representing a point in the scene. 
In this way, Gaussian Splatting represents the function $\phi_{\theta}(x, v)$ as a set of 3D Gaussians.

Each Gaussian is defined by its mean $\mu$, covariance matrix $\Sigma$, opacity $\delta$, and color $c$.
By representing the color with spherical harmonics, view-dependent effects can also be captured.
Since the covariance matrix $\Sigma$ has to be positive semi-definite, which is difficult to enforce during optimization, the covariance matrix is obtained by $RSS^TR^T$, where $S$ is a scaling matrix and $R$ a rotation matrix obtained from a quaternion.
The covariance matrices of the Gaussians are initialized accounting for their neighbors to conservatively cover the surfaces and prevent holes in the reconstruction.

The optimization process used in Gaussian Splatting is similar to the one used in NeRFs, where volume rendering is used to generate images by querying density and color along the rays emanating from the camera.
However, due to the sparse nature of the representation achieved with Gaussian Splatting, the rendering process can be efficiently computed by projecting the Gaussians into the image and combining them using alpha blending.
Furthermore, since the initial set of Gaussians may not be sufficient to capture all the necessary geometric and color details, the Gaussians are periodically split or cloned based on their accumulated $\mu$ gradient.
However, this splitting is designed to represent the original scene and might not be sufficient for its stylized version.
As we discuss in the upcoming sections, we propose a fast, high-quality, and consistent approach for stylizing 3D scenes by modifying the style and geometry of scenes represented by Gaussian Splatting with the style of an additional image or texture.

\section{Methodology of $\mathscr{G}$-Style: Gaussian Splatting with Style}
In this section, we describe our proposed algorithm: $\mathscr{G}$-Style.
We first provide an overview of our approach (also illustrated in Figure~\ref{fig:overview}), followed by a detailed explanation of its substeps.

\vspace{-5pt}
\subsection{Overview}
Our algorithm takes as input a scene represented with a set of Gaussians $\mathcal{G}$ and a set of \rev{ground truth} images $\mathcal{I}_{gt}$. 
Subsequently, it modifies $\mathcal{G}$ based on the style provided from a style exemplar $\mathcal{I}_s$.
First, we \textit{pre-process} $\mathcal{G}$ to remove long narrow Gaussians and Gaussians covering large areas, making the initial set of Gaussians uniform.
Once the initial representation has been pre-processed, we create an additional color $c_s$ associated with each Gaussian which is initialized with the original color $c_{gt}$.
These new colors $c_s$ are modified during a \textit{stylization} process that uses a composition of several losses to preserve different properties of the style of $\mathcal{I}_s$.
Since the geometry provided by the initial pre-processing step can be limited to represent detailed style features, $\mathcal{G}$ undergoes a \textit{geometric fine-tuning} step.
In this step, the Gaussians are split based on the gradient of $c_s$ and fine-tuned to match the original scene images $\mathcal{I}_{gt}$ by modifying $\mu$, $\Sigma$, $\delta$, and $c_{gt}$.
The stylization and geometric fine-tuning steps are repeated until convergence.

\vspace{-5pt}
\subsection{Pre-processing Step}
\rev{
Although Gaussian Splatting offers high-quality scene representation, the original approach has limitations. 
It often generates large flat Gaussians to depict uniform, flat surfaces (e.g., walls), and narrow elongated Gaussians to capture high-frequency details. 
The latter might also come as a byproduct of the optimization process.}
When incorporating an additional style into the scene, large flat areas might need additional Gaussians to incorporate extra details, while \rev{already highly detailed areas may not need more Gaussians to stylize them properly.}
Therefore, in the initial step of our approach, the scene undergoes a Gaussian normalization process where the resulting representation $\mathcal{G}$ is composed of Gaussians of similar size and shape.

\noindent \textbf{Flat Gaussian Split.}
To detect under-sampled areas, we compute the approximated maximum projected area of each Gaussian, $A_i$.
We compute $A_i$ by multiplying the two highest components of the Gaussian's scaling matrix $S_i$.
Then, we mark for splitting all Gaussian above a threshold $t_f = \mu_A + \gamma\sigma_A$, where $\mu_A$ is the mean of all $A$ in the scene, $\sigma_A$ is the standard deviation of $A$, and $\gamma$ is a user-defined parameter controlling the number of Gaussian mark for splitting.
As splitting Gaussians modifies the geometry of the scene, we are obliged to optimize them. 
For this, we employ the same optimization algorithm as in the original work of Kerbl et al.~\cite{kerbl3Dgaussians}.

\begin{figure}[t!]

  \includegraphics[width=\linewidth]{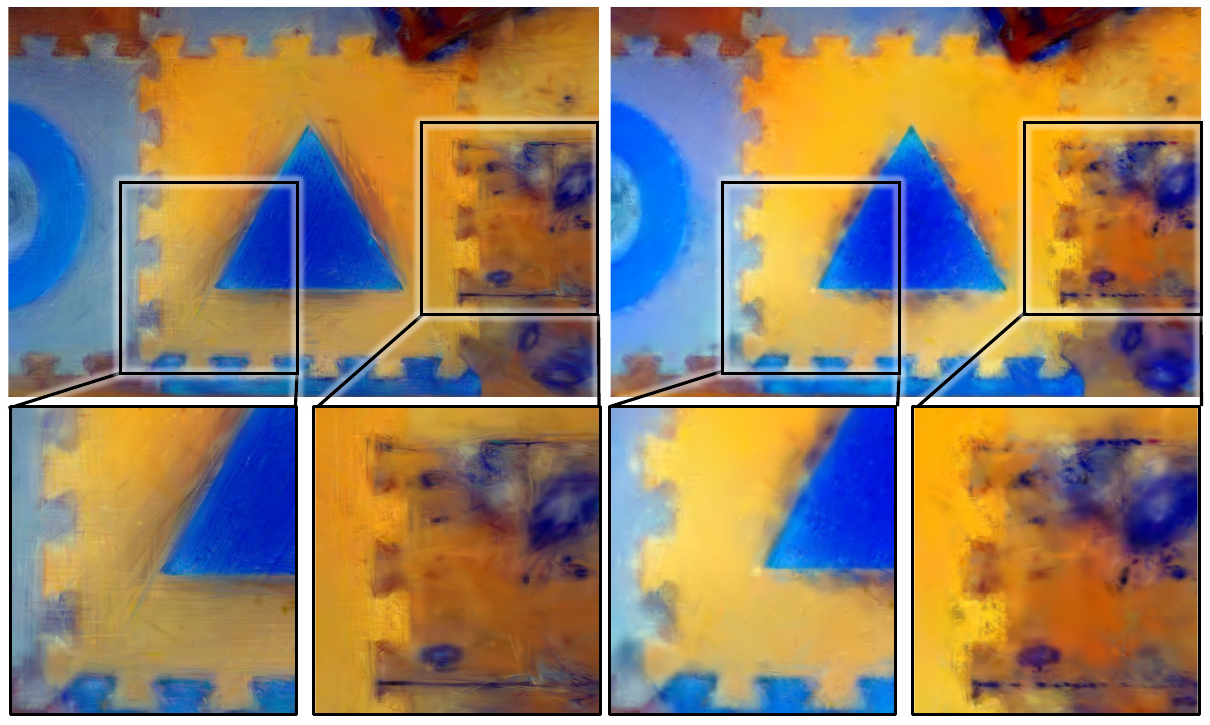}
  \caption{\textbf{Gaussian Normalization: } We normalize (\rev{right}) the size of narrow Gaussians to avoid multiple overlying Gaussians (\rev{left}).
  }
  \label{fig:compaction}
\end{figure}

\noindent \textbf{Narrow Gaussian Normalization. }
To identify Gaussians with a narrow shape, we compute their elongation factor $E_i$ by dividing the highest component of each Gaussian's scaling matrix $S_i$ by its second highest component.
If $E_i$ is above a certain user-defined threshold $t_e$, we mark it for normalization, which sets the largest component to the average of the largest and second-largest components.
\rev{We repeatedly perform this operation during the optimization process after the flat Gaussian split.
This optimization process retrains the scene to match the appearance of the ground truth images $\mathcal{I}_{gt}$.
As such, it corrects the deformations caused by this narrowing step while keeping the Gaussians more rounded.}
In Figure~\ref{fig:compaction}, we illustrate the effect of normalizing Gaussians (\rev{right}) as opposed to not normalizing them (left). 

\noindent \textbf{Diffuse Color Transform. }
In 3D style transfer, where there is no concrete ground truth image associated with a view, enforcing multi-view consistency is key for seamless navigation through the scene.
In the original Gaussian Splatting algorithm, spherical harmonics enable modeling view-dependent effects such as reflections. 
However, they also generate undesirable artifacts on the stylized version of the scene, where each view direction could result in a completely different stylization.
To avoid this, we only use the zeroth term of the spherical harmonics representation, limiting the model to represent diffuse objects only, therefore, enforcing view consistency.

\begin{figure}[t!]
  \includegraphics[width=0.5\linewidth]{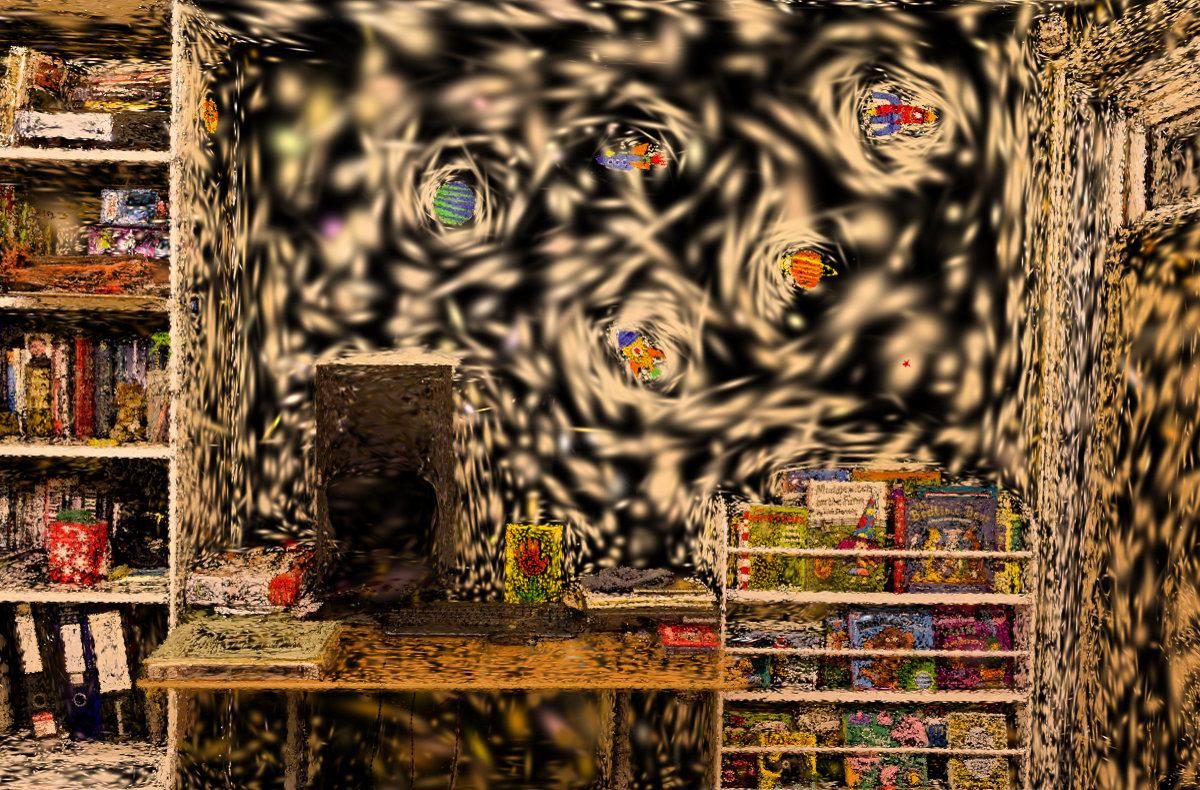}
  \includegraphics[width=0.5\linewidth]{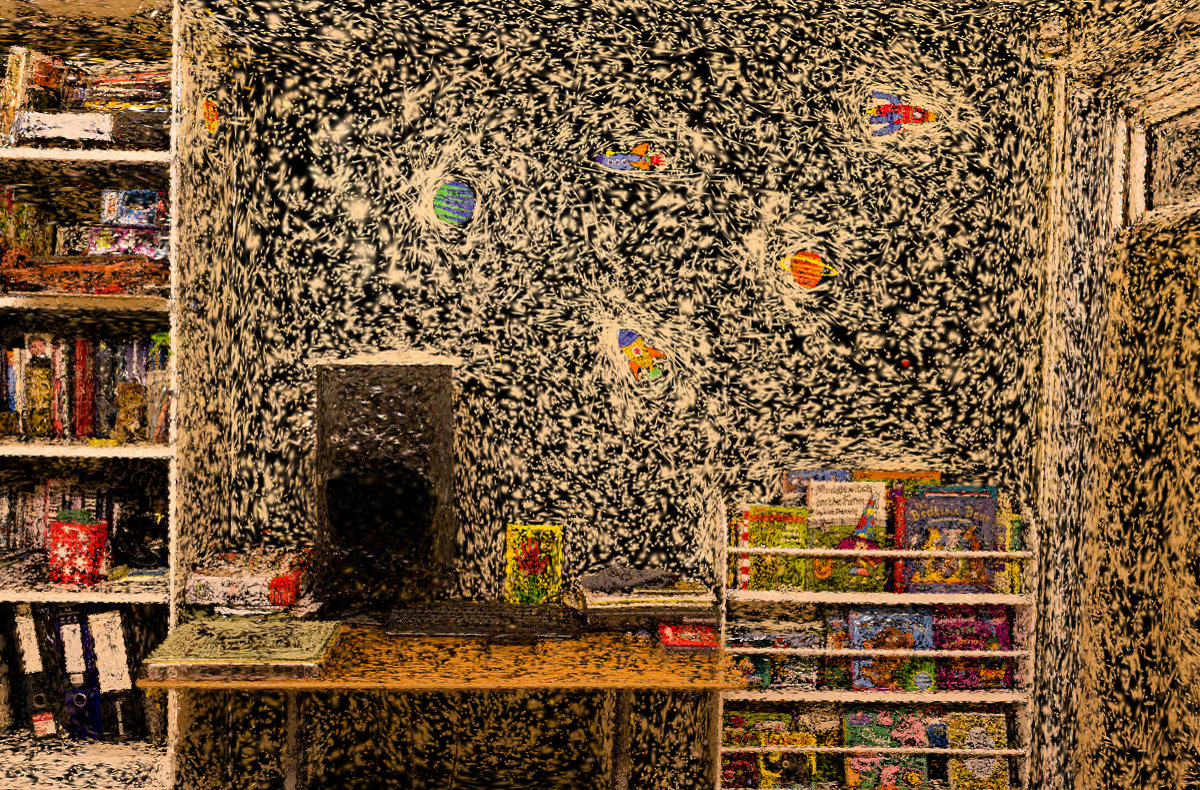}
  \caption{\textbf{Pre-processing: }The effect of our pretraining step. Left: before pretraining, right: after pretraining. The scale of Gaussians is set to 0.25, otherwise both images would look identical.}
  \label{fig:pretraining}
\end{figure}

\noindent \textbf{Parameterizations. }
We perform five rounds of the splitting--normalization--optimization process, which in our experiments offers a good balance between the pre-computation time and the even distribution of Gaussian sizes.
Initially, we set $\gamma$ to 1.1, but in each subsequent round, we multiply it by 1.125 to deal with very large Gaussians that may still be remaining.
The initial value for $\gamma$ is quite low and, thus, we split at most 5\% of the largest Gaussian in each round.
Additionally, we set the threshold for elongation $t_e$ to 1.5, which allows a degree of elongation but normalizes the shape of those Gaussians that heavily affect the final appearance.
After pre-processing, the resulting set of Gaussians still matches the visual characteristics of a scene, but the distribution of their perceived sizes is uniform, as can be seen in Figure~\ref{fig:pretraining}.
Note that incorporating more Gaussians into the scene results in additional details---and additional memory demands.
However, since we limit our representation to diffuse objects, the storage required to store the color of each Gaussian is reduced from $192$ bytes to $12$ bytes for the three components of the diffuse color.
Our implementation, thus, allows us to increase the number of Gaussians to represent the scene while at the same time reducing memory consumption.

\subsection{Stylization Step}

Once we have pre-processed the scene, we start the stylization process.
During stylization, we render the scene from multiple views using $c_s$ and compute several losses over the images carefully designed to preserve the style of $\mathcal{I}_s$ at different scales.
We hereby describe in detail the different losses used in our algorithm.

\noindent \textbf{Low-frequency Style. }
The style of an image is determined by color and high-frequency details and also by large-scale patterns and features.
To preserve these low-frequency features of the style image, we employ a CLIP-based~\cite{radford2021learning} loss.
We encode our rendered images $\mathcal{I}^{k}_r$ and the style image $\mathcal{I}_s$ into a feature vector using the image encoder of the CLIP model, $\mathcal{C}$.
Then, our loss is defined as the similarity between these feature vectors where similarity is measured as the $l_2$:
\begin{equation}
    \mathcal{L}_{CLIP}=\frac{1}{K}\sum_{k=1}^K \lVert \mathcal{C}(\mathcal{I}^{k}_r) - \mathcal{C}(\mathcal{I}_s) \rVert^2_2
\end{equation}
where $K$ is the number of rendered images in the batch.

\noindent \textbf{High-frequency Style. }
Our CLIP-based loss can capture large-scale patterns in the style image. 
Yet, fine details, such as brush strokes in a painting, might not be well represented with this loss.
Therefore, following Zhang et al.~\cite{zhang2022arf}, we use a nearest neighbor feature matching (NNFM) loss utilizing a pre-trained VGG-16 network~\cite{simonyan2014vgg16} to capture the high-frequency style patterns.
\rev{The architecture of a feature extractor can play a significant role in the quality of the generated results. We selected VGG-16 for its proven effectiveness in previous style transfer solutions, facilitating easier comparisons with those methods.}
The NNFM loss is a replacement for the widely used Gram matrix-based loss of Gatys et al.~\cite{gatys2015styletransfer}.
Instead of matching statistics of feature maps $\mathcal{F}_{r}$ and $\mathcal{F}_{s}$, this loss searches for nearest neighbors in the feature space.
Let $\mathcal{F}_{r}$ and $\mathcal{F}_{s}$ be the VGG-16 features maps of $\mathcal{I}_r$ and $\mathcal{I}_s$ respectively, and let $\mathcal{F}(i, j)$ be the feature vector at pixel location $(i, j)$. 
The NNFM loss is then defined as:
\begin{equation}
    \mathcal{L}_{NNFM}(\mathcal{F}_{r}, \mathcal{F}_{s}) = \frac{1}{N}\sum_{i,j}\min_{i',j'} \phantom{i}D(\mathcal{F}_{r}(i, j), \mathcal{F}_{s}(i',j'))
\end{equation}
where $N$ is the number of pixels in $\mathcal{F}_{r}$, and $D$ is the cosine distance between two vectors.
As in Zhang et al.~\cite{zhang2022arf}, we use feature maps from the third block of VGG-16.

\noindent \textbf{Regularization. }
To avoid the deterioration of features that are necessary for scene understanding, like in the work of Zhang et al.~\cite{zhang2022arf}, we utilize a content loss $\mathcal{L}_C$, which is the $l_2$ loss between the features of rendered images $\mathcal{F}_{r}$ and ground truth images $\mathcal{F}_{gt}$.
This loss also uses the features from the third block of VGG-16.
Additionally, we incorporate a total variation term in our loss $\mathcal{L}_{TV}$ to prevent noise in the resulting renderings.

\noindent \textbf{Complete Style Loss. }
Our final loss is a weighted sum of all the losses, given by the equation:
\begin{equation}
    \mathcal{L} = \lambda_{CLIP}\mathcal{L}_{CLIP} + \lambda_{NNFM}\mathcal{L}_{NNFM} + \lambda_{C}\mathcal{L}_{C} + \lambda_{TV}\mathcal{L}_{TV}
\end{equation}

We perform the stylization process for 15 epochs. Note that forward-facing scenes do not have as strict multi-view consistency requirements as 360° scenes due to the limited viewing angles of the ground truth images. As such, they converge more easily compared to 360° scenes. To account for the differences between these two types of scenes, we use different parameterizations. For forward-facing scenes, we use an exponentially decaying learning rate from 1e-1 to 1e-2, and for 360° scenes the learning rate decays from 1e-2 to 5e-3. Also, we set $\lambda_{CLIP}$ to 10, $\lambda_{NNFM}$ to 100, $\lambda_C$ to 0.05, and $\lambda_{TV}$ to 1e-4. For 360° scenes, we set $\lambda_{NNFM}$ to 10.

\subsection{Geometric Fine-tuning Step}

In the pre-processing step, we place more Gaussians in the undersampled areas, making the Gaussians more uniform in size.
This step is conservative enough to not overly increase the number of Gaussians and, thus, not oversample a given scene.
With the new Gaussians, it is possible to synthesize features that otherwise could not be represented. 
However, the size of the Gaussians still limits the synthesis of very fine features in our stylized scene.

To overcome this, during the stylization process, we periodically split Gaussians based on their $c_s$ gradient. 
Similarly to the work of Kerbl et al.~\cite{kerbl3Dgaussians}, we keep an accumulation buffer $\mathcal{B}$ to store the norm of the gradient $c_s$. 
After each iteration and for each Gaussian, we add the norm of the $c_s$ gradient to $\mathcal{B}$ and periodically split a user-defined percentage of Gaussians with the highest value.
Since splitting Gaussians modifies the geometry of the scene, after each splitting, we optimize their $\mu$, $\Sigma$, $\delta$, and $c_{gt}$ in the same optimization process as Kerbl et al.~\cite{kerbl3Dgaussians}. 

\subsection{Style Color Matching }
To ensure a similar color distribution between the resulting stylized 3D scene and the original style image, at the beginning of our algorithm, we match the mean and covariance matrix of colors from our ground truth images $\mathcal{I}_{gt}$ and the style image $\mathcal{I}_s$.
Let $C$ be a matrix containing pixel colors of the ground truth images of the scene $\mathcal{I}_{gt}$ and $S$ be a matrix of pixels from $\mathcal{I}_s$, where each row is one pixel and columns are used for RGB components.
We analytically solve for a linear transformation $A$, such that $E(AC)=E(S)$ and \rev{$Cov(AC)=Cov(S)$}, and finally modify our initial Gaussian Splatting scene representation so that the rendered color images match with the color-corrected ground truth images.
Even though this color transfer step assumes unimodal distributions of colors, in our experiments, we empirically identified that it is sufficient for all tested style images.
\rev{Since optimizing with a loss that considers the activations of hidden layers does not guarantee that the resulting colors are accurate to the style, we apply the same correction at the end of our algorithm to $\mathcal{I}_{r}$.}
\begin{figure*}
\centering
  \includegraphics[width=.975\linewidth]{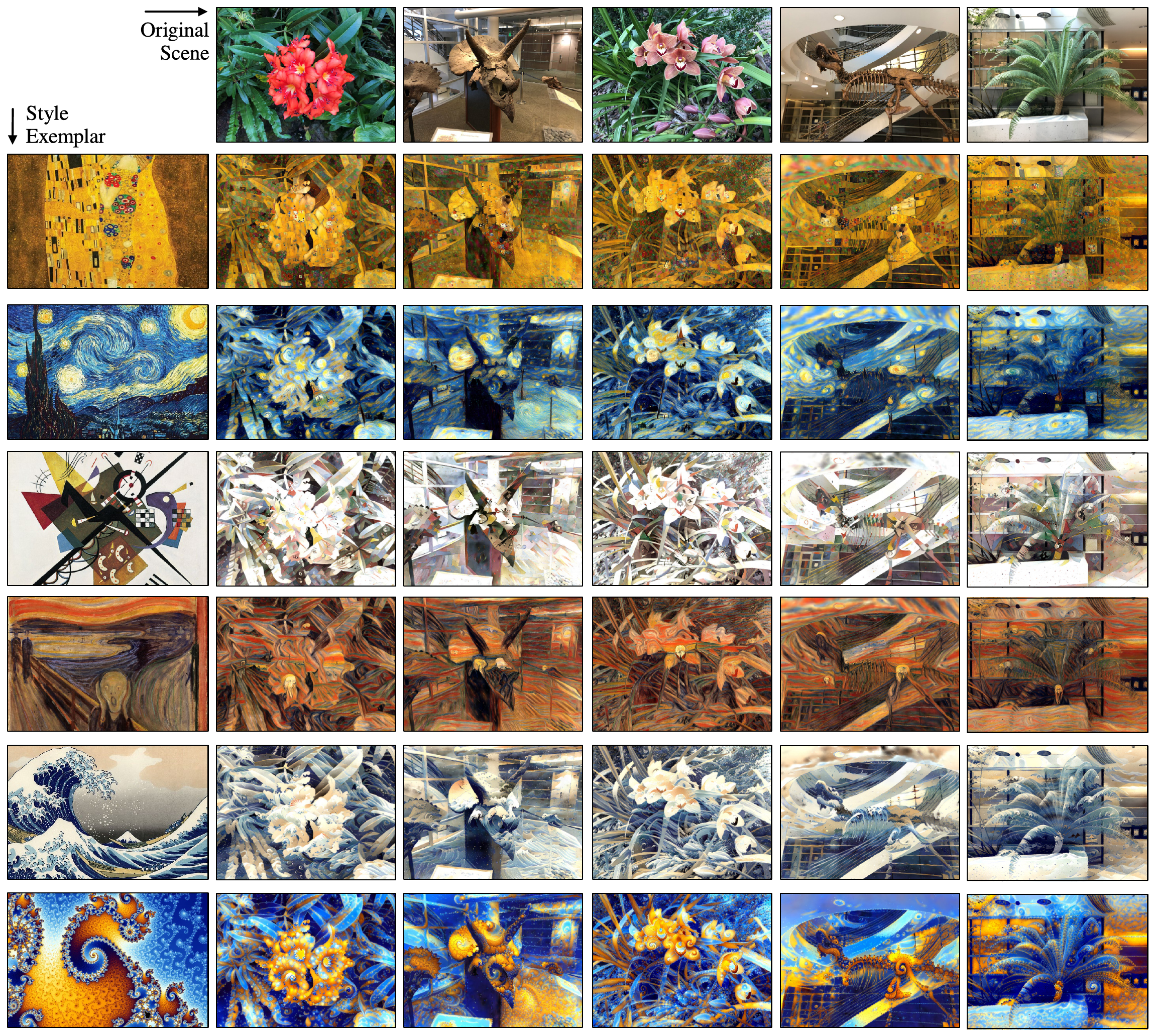}
     \vspace{-10pt}
  \caption{Results generated with our approach, $\mathscr{G}$-Style, for five forward-facing scenes (columns) given six style exemplars (rows).    \vspace{-10pt}
  }
  \label{fig:results_forwardOurs}
\end{figure*}

\begin{figure*}
\centering
  \includegraphics[width=.975\linewidth]{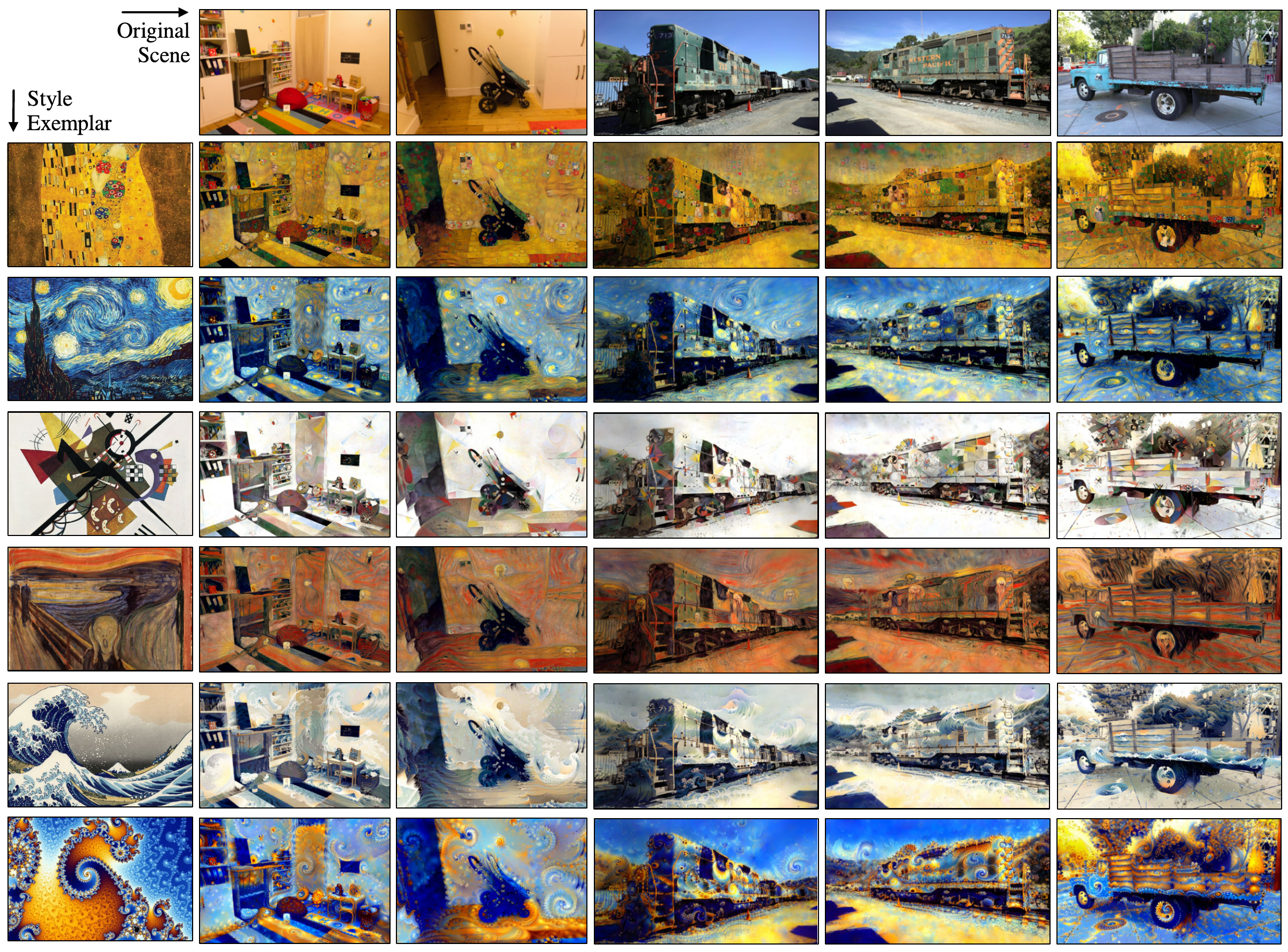}
  \vspace{-10pt}
  \caption{Results generated with our approach, $\mathscr{G}$-Style, for five 360° scenes (columns) given six style exemplars (rows).   
  }
  \label{fig:results_360Ours}
\end{figure*}

\section{Results}
In this section, we provide an analysis of the results produced by our method.
Additionally, we offer a comparison to other leading state-of-the-art approaches.
All of our results are generated using a modified 3D Gaussian Splatting codebase~\cite{3dgs_code}, \rev{which is publicly available in our GitHub Repository (\url{https://github.com/AronKovacs/g-style})}.

\subsection{Datasets}

To evaluate our approach, we prepared a series of forward-facing scenes: \textit{Flower}, \textit{Horns}, \textit{Orchid}, \textit{T-Rex}, and \textit{Fern} (employed in the work of Mildenhall et al.~\cite{mildenhall2019llff}) and 360° scenes: \textit{Playroom}~\cite{hedman2018deep}, and \textit{Truck} and \textit{Train}~\cite{Knapitsch2017} together with a variety of styles ranging from classical paintings to abstract images: \textit{The Starry Night} by \textit{Vincent van Gogh}, \textit{The Scream} by \textit{Edvard Munch}, \textit{The Great Wave off Kanagawa} by \textit{Hokusai}, \textit{On White II} by \textit{Wassily Kandinsky}, \textit{The Kiss} by \textit{Gustav Klimt}, and a colored image of the \textit{Mandelbrot Set} created by \href{https://commons.wikimedia.org/wiki/User:Wolfgangbeyer}{\textcolor{blue}{Wolfgang Beyer}} (\href{https://creativecommons.org/licenses/by-sa/3.0/}{\textcolor{blue}{\emph{CC BY-SA}}}).
The scenes contain vastly different complexities in the represented stimuli---including highly detailed areas, such as the bookcases in the \textit{Playroom}, but also flat white walls.
Our generated results for the forward-facing scenes can be seen in Figure~\ref{fig:results_forwardOurs} and for the 360° scenes in Figure~\ref{fig:results_360Ours}.

\subsection{Comparison to the State of the Art}

We compare our approach to three state-of-the-art approaches: a recent NeRF-based approach, Artistic Radiance Fields (ARF)~\cite{zhang2022arf}, a recent zero-shot style transfer method for NeRF scenes, StyleRF~\cite{liu2023stylerf}, and a recent pre-print that performs style transfer for Gaussian Splatting scenes, StyleGaussian~\cite{liu2023stylegaussian}. 
For this comparison, we used their official implementations and pre-trained checkpoints~\cite{arf_code, stylerf_code, liu2023stylegaussian}.
\rev{We do not compare against Gaussian Splatting in Style~\cite{saroha2024gaussian} and StylizedGS~\cite{zhang2024stylizedgs} as their implementations are not publicly available at the time of writing this paper.}

ARF uses the NNFM loss to transfer artistic style and also utilizes explicit color matching to achieve more accurate colors. 
This approach can use different volumetric scene representations as its backbone.
However, the only publicly available implementation~\cite{stylerf_code} uses TensoRF~\cite{Chen2022ECCV}, which in essence is a 3D voxel-based representation, decomposed into several lower-rank tensors. 
Other representations showcased in the original paper are neural radiance fields and Plenoxels~\cite{yu2022plenoxels}, but these could not be tested as they were not publicly available.
StyleRF also uses TensoRF as its backbone. 
This approach is based on embedding high-dimensional features into the structure of the scene and, during rendering, uses them to transfer style.
StyleGaussian is similar to StyleRF but uses Gaussian Splats as its backbone. 

\begin{figure*}
\centering
  \includegraphics[width=1\linewidth]{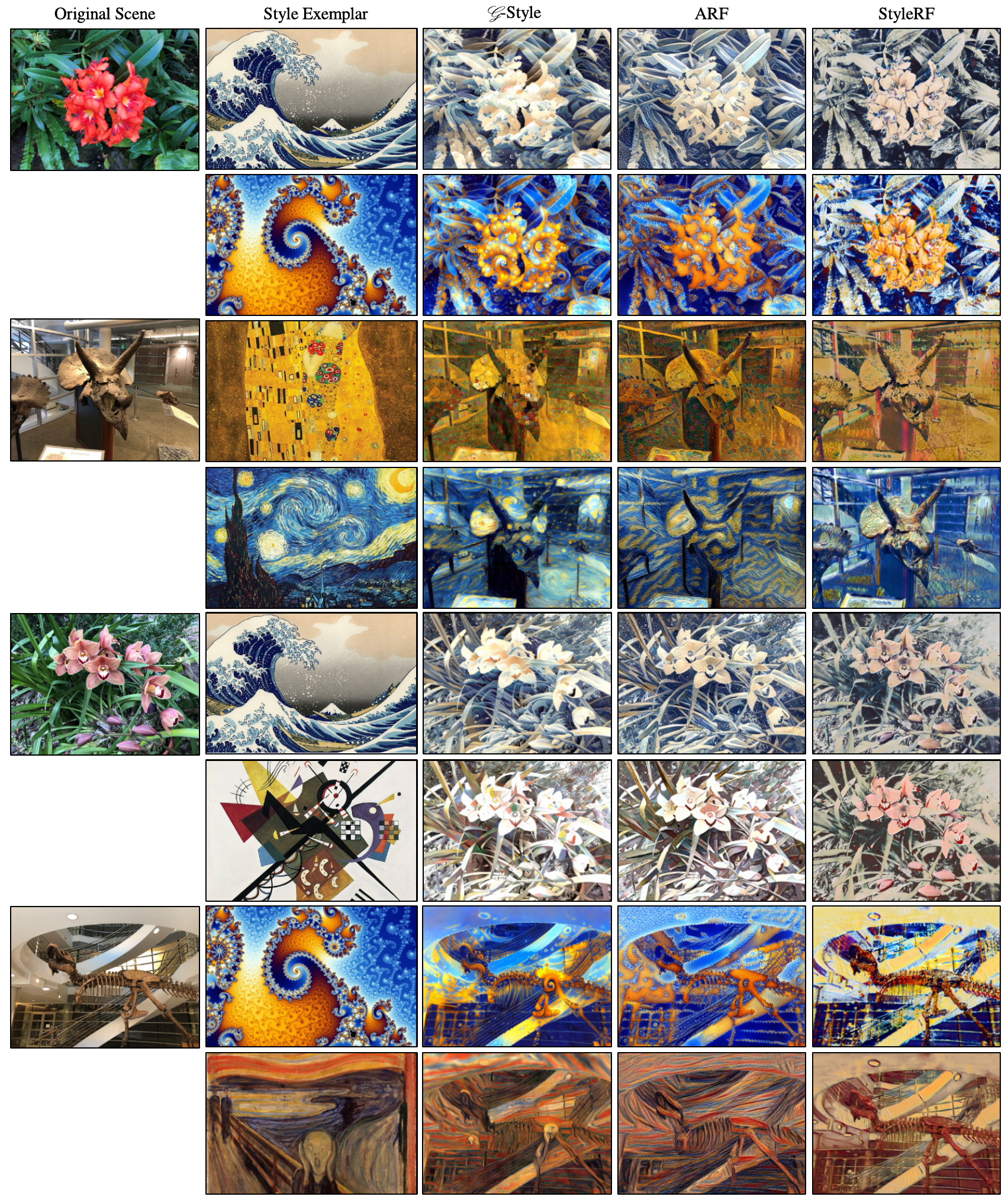}
     \vspace{-10pt}
  \caption{Results generated with our approach ($\mathscr{G}$-Style), ARF~\cite{zhang2022arf}, and StyleRF~\cite{liu2023stylerf} (columns) for four forward-facing scenes and two style exemplars for each scene (rows).    \vspace{-10pt}
  }
  \label{fig:results_forward}
\end{figure*}

\begin{figure*}
\centering
  \includegraphics[width=1\linewidth]{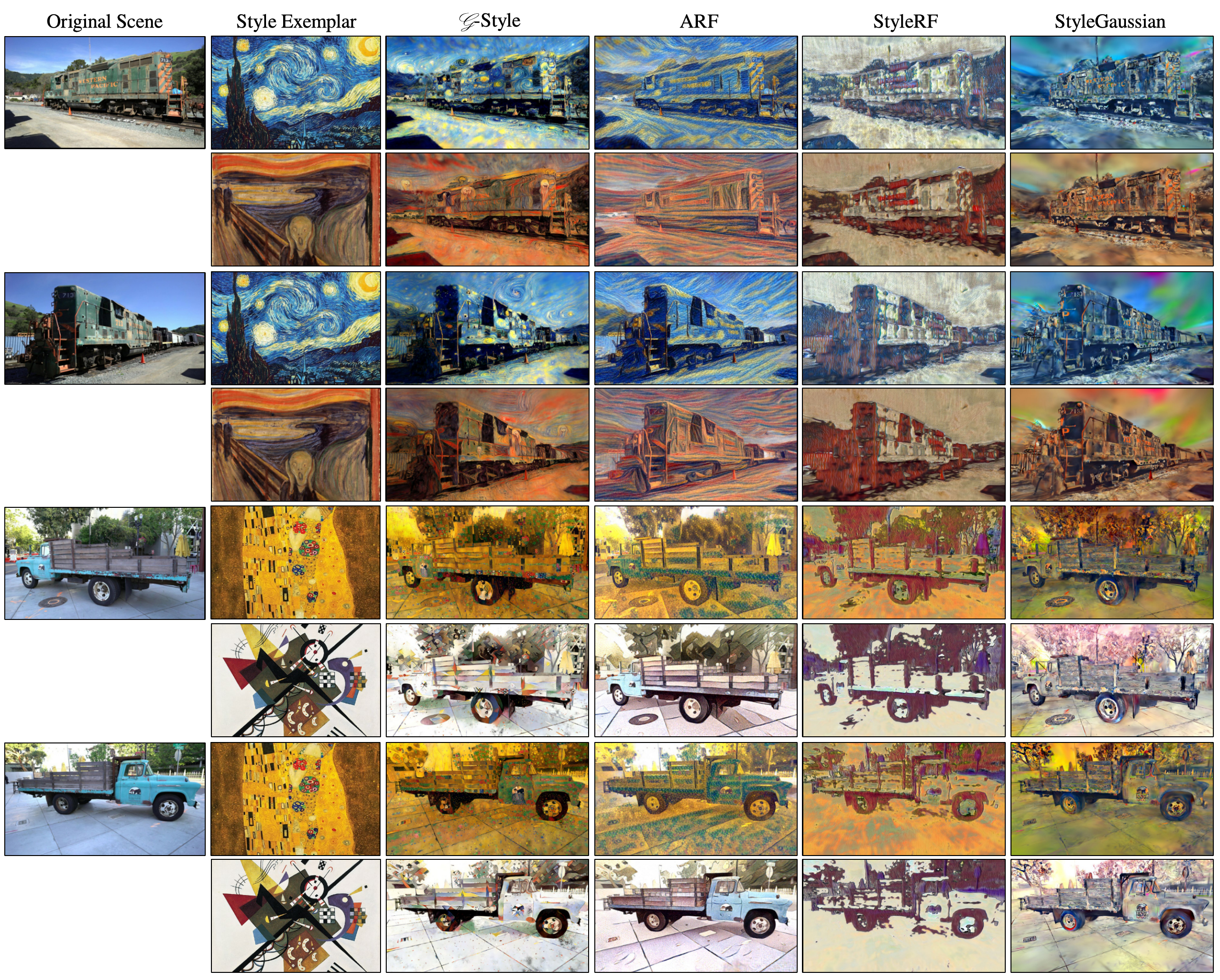}
     \vspace{-10pt}
  \caption{Results generated with our approach ($\mathscr{G}$-Style), ARF~\cite{zhang2022arf}, StyleRF~\cite{liu2023stylerf}, and StyleGaussian~\cite{liu2023stylegaussian} (columns) for 360° scenes and two style exemplars for each scene (rows).    \vspace{-10pt}
  }
  \label{fig:results_360}
\end{figure*}

\noindent \textbf{Qualitative Comparison.}
A comparison for the forward-facing scenes can be seen in Figure~\ref{fig:results_forward} and for the 360° scenes in Figure~\ref{fig:results_360}.
In these figures, we present the comparison of our method to the available checkpoints of other approaches for different scenes.
Our approach exhibits better or, at least, comparable results to the other methods.
Both StyleRF and StyleGaussian do not faithfully capture the visual style, because they are not able to synthesize small patterns (as is the case for the \textit{Mandelbrot} style in Figure~\ref{fig:results_forward}, second and seventh row) nor brushstrokes (for \textit{The Starry Night} and \textit{The Scream} styles) in Figures~\ref{fig:results_forward} and~\ref{fig:results_360}. 
These two approaches also fail to recreate any bigger patterns, such as in the two \textit{Truck} scenes of Figure~\ref{fig:results_360}, especially in the stylization with \textit{On White II}.
Furthermore, StyleGaussian maintains the original Gaussians as resulting from the reconstruction phase. 
It is, thus, not able to create any meaningful patterns in undersampled areas, which can be easily seen on the walls in the examples of Figure~\ref{fig:results_forward}, and on the ground or in the sky in all the examples depicted in Figure~\ref{fig:results_360}.
Moreover, as shown in the \textit{Train} scene in Figure~\ref{fig:results_360}, StyleGaussian produces large colorful Gaussians in the sky which do not match any of the used style images.
Thus, for the scenes and styles we chose, StyleRF and StyleGaussian cannot reliably generate style-specific details and patterns; but rather focus on recoloring the scenes.
Yet, the visual style of an image goes beyond just the colors.

Similarly to ARF, we produce high-frequency details, but by utilizing the CLIP loss, we can also create bigger patterns.
This is something that ARF seems to struggle with.
In the case of \textit{The Starry Night} (in both Figure~\ref{fig:results_forward} and~\ref{fig:results_360}), our stylized scenes contain brush-like patterns, but also moon-like and star-like shapes not present in the results of ARF.
When using \textit{The Scream} our method occasionally also creates head-like shapes (see the femur of the \textit{T-Rex} in Figure~\ref{fig:results_forward} and the surface of the \textit{Train} in Figure~\ref{fig:results_360}).
If this is not desired, it could be removed by modifying the style image and/or cropping it.
Furthermore, \textit{On White II} consists of simple patterns using only a single color.
While our method can capture those patterns and create colorful shapes, ARF only produces desaturated areas, which are not truthful to the painting.
This is evident in the sixth row of Figure~\ref{fig:results_forward} and the \textit{Truck} in Figure~\ref{fig:results_360}.
Finally, when using \textit{The Kiss}, the flower pattern in Figure~\ref{fig:results_forward} is less blurry and more recognizable with our method, as opposed to ARF.
We conclude that, while ARF can produce highly stylized images that match certain style images, it focuses only on fine details and ignores bigger patterns which may be important to capture a given style successfully.

\begin{figure}
\centering
  \includegraphics[width=\linewidth, height=2.8in]{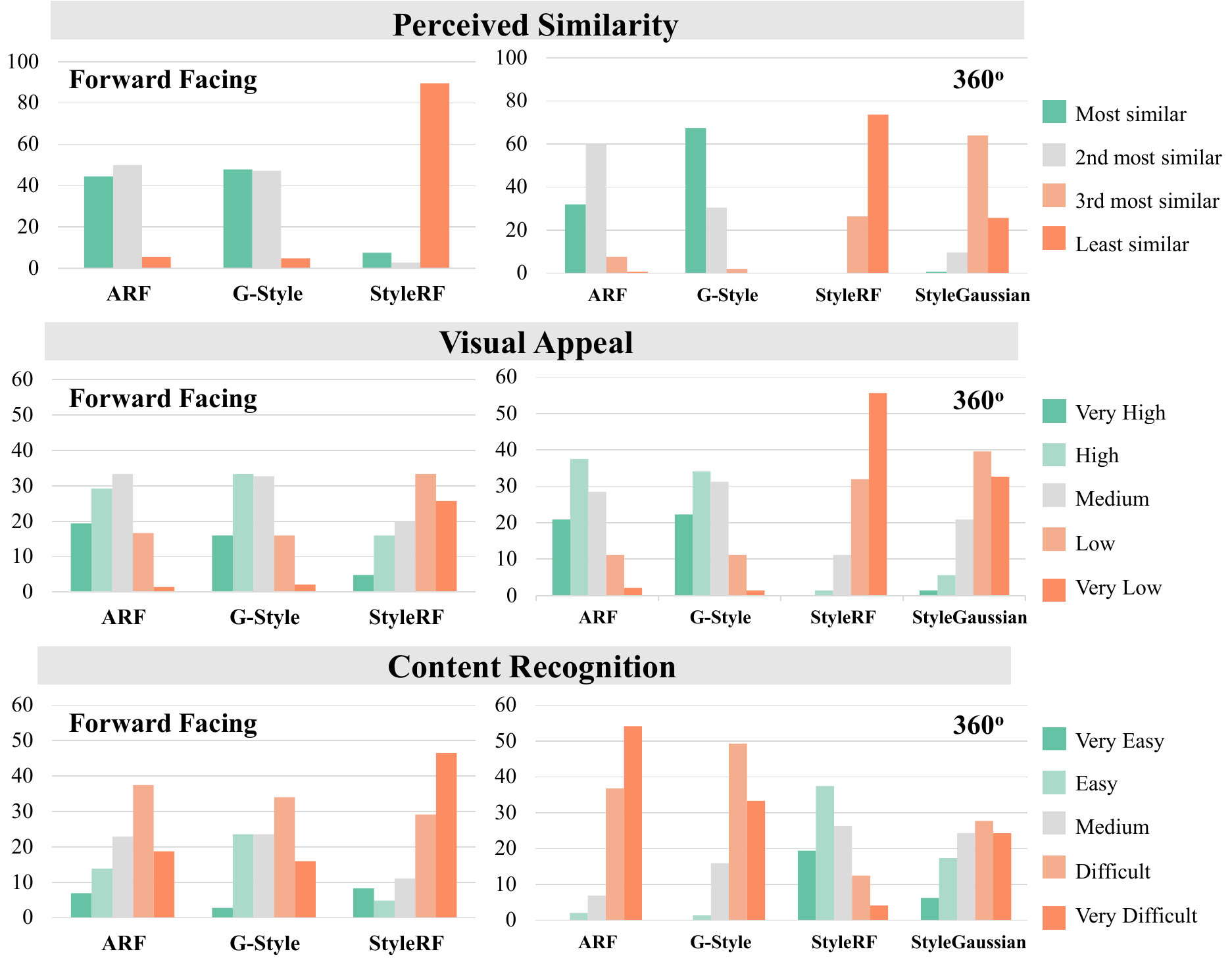}    \vspace{-10pt}
  \caption{User study results: our approach ($\mathscr{G}$-Style) vs. ARF~\cite{zhang2022arf}, StyleRF~\cite{liu2023stylerf}, and StyleGaussian~\cite{liu2023stylegaussian}.    \vspace{-10pt}
  }
  \label{fig:outcome_study}
\end{figure}

\noindent \textbf{Speed Comparison.}
We measured the optimization times of all the compared methods, and we performed all of our measurements on an NVIDIA L4 in Google Colab. The chosen approaches are fundamentally different: our approach and ARF~\cite{arf_code} optimize directly the colors of a scene, while StyleRF~\cite{liu2023stylerf} and StyleGaussian~\cite{liu2023stylegaussian} embed VGG features in the scene and later use them during rendering to stylize them.
For the tested forward-facing scenes, our pre-processing step takes approximately 5 minutes, and stylization 3--8 minutes depending on the complexity of the scene, the number of ground truth images, and the size of the style image. For the same scenes, ARF needs 2--7 minutes. For the tested 360° scenes, the preprocessing step of our method takes 8 minutes, and stylization 20--28 minutes. For the same scenes, ARF requires 23--33 minutes. Note that depending on the particular scene and the quality of its reconstruction, we may not need to perform the pre-processing step, as its purpose is to ensure an approximately uniform distribution of the shapes of Gaussians.
For StyleGaussian, the process when the embedded features are infused with the style information takes approximately 18 hours per scene, and the rendering of stylized images can be done in real-time. StyleRF needs 30--36 seconds per frame, which includes both style transfer and rendering.

\subsection{User Study}
To evaluate our method, we conducted an informal, online user study with 24 participants, where we used several of the cases shown in Figures~\ref{fig:results_forwardOurs}--\ref{fig:results_360}.
We presented each participant with the results of our approach, ARF, StyleRF, and StyleGaussian together with the respective style exemplars and original scenes.
Without disclosing any information about any of the approaches, we interviewed the participants to gain feedback about the outputs.
Specifically, we asked them to rank the approaches w.r.t. their similarity to the provided style exemplar. 
For each of the generated results, we also asked the study participants to rate their visual appeal and ability to recognize the original scene content on a 1--5 Likert scale.  

The analyzed outcomes of the user study are shown in Figure~\ref{fig:outcome_study}.
The study participants ranked our approach as most similar to the style exemplar ($47.9$\% of the participants for the forward-facing scenes vs. $67.4$\% for the 360° scenes), followed by ARF ($44.4$\% for the forward-facing scenes vs. $32$\% for the 360° scenes). 
StyleRF ($7.6$\% for the forward-facing scenes vs. $0$\% for the 360° scenes) and StyleGaussian ($0.7$\% for the 360° scenes) ranked last. 
The differences between our approach and ARF are statistically significant only for the 360° scenes, as shown with an ANOVA test followed by pairwise $t$-tests.
The most visually appealing approaches are deemed to be ours and ARF, as indicated visually in the plots of Figure~\ref{fig:outcome_study}.
However, the distributions of the ratings of our approach and ARF do not statistically significantly differ.
In terms of content recognition, our approach and ARF are comparable to each other. 
Both are better than StyleRF for forward-facing scenes, but this changes for the 360° scenes, where the ratings of StyleRF (followed by StyleGaussian) are higher.
This is to be expected as the effect of the latter two on the stylization of the scenes is less drastic than the former. 
To sum up, according to our study participants, our approach is comparable to ARF in all investigated aspects---yet, for 360° scenes, ours yields results more faithful to the style.

\begin{figure*}
  \centering
  \includegraphics[width=\linewidth]{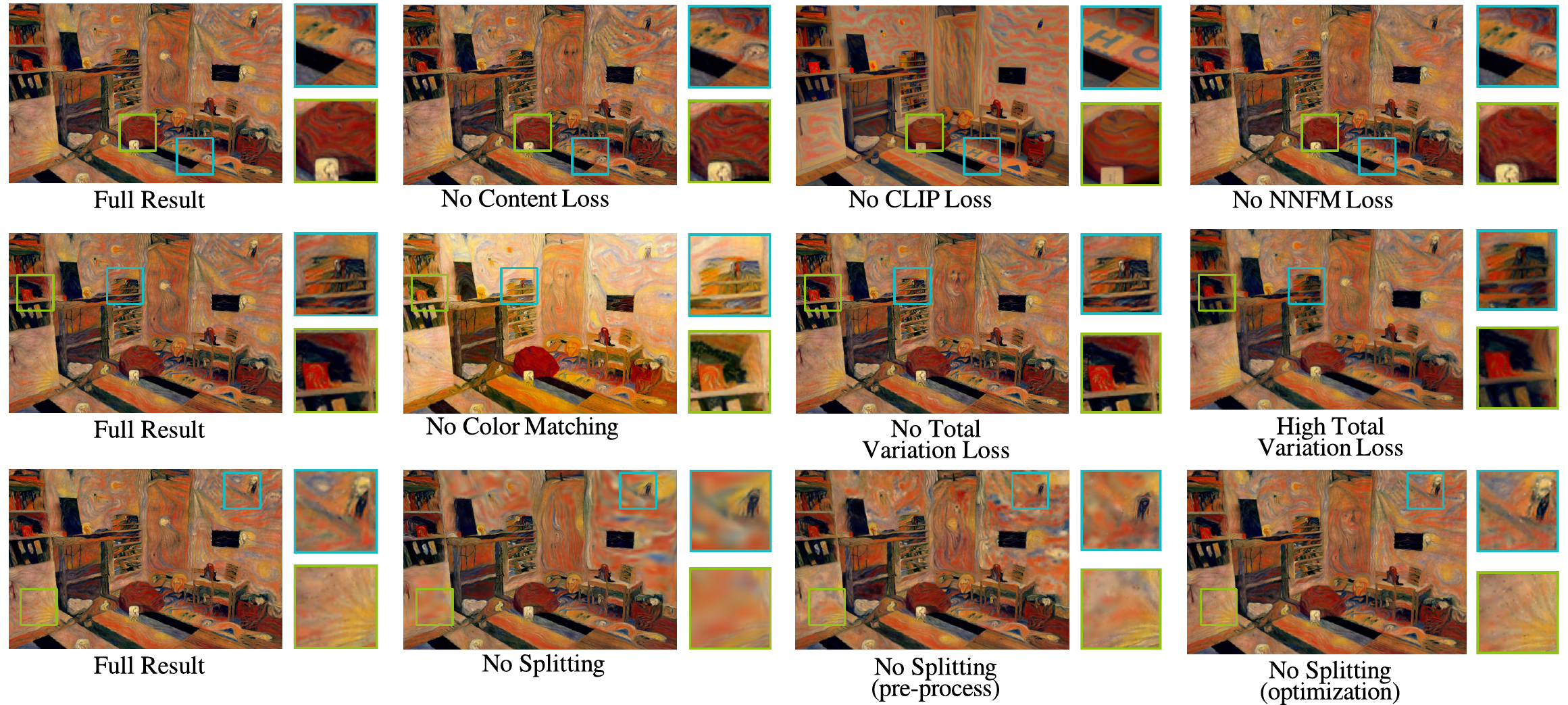}
     \vspace{-10pt}
  \caption{Ablation studies performed for our approach.   
  }
  \label{fig:ablations}
\end{figure*}

\subsection{Ablations}

Our loss does not enforce color transfer and relies on pre-trained networks to synthesize new features. 
Often, these new features are patterns that do not necessarily have the same color as in the style image. 
\rev{When we do not perform any explicit \textbf{color matching}---both during and after training---synthesized images are discolored (see Figure~\ref{fig:ablations}).}
Conversely, with color matching, the colors are truthful to the style image, as also shown in the work of Zhang et al.~\cite{zhang2022arf}.
Introducing another style loss, $\mathcal{L}_{CLIP}$, still does not ensure that the optimization process matches the color statistic in synthesized images.
\textbf{Without splitting}, the number and shape of Gaussians are the same as after the initial optimization with the ground truth images. 
This results in blurred areas, where not many Gaussians were originally needed, such as the walls marked in Figure~\ref{fig:ablations}. By splitting Gaussians, we are able to introduce details even in those areas.
The \textbf{content loss} conditions the stylization. 
This loss helps to keep certain features intact or recognizable, e.g., the letter on the floor in the \textit{Playroom} scene (see Figure~\ref{fig:ablations}). Furthermore, it helps to suppress noise, which is not present in the ground truth images.
Moreover, the \textbf{CLIP loss} enables our method to focus on bigger features. 
Without it, only very fine features are transferred. 
For example, without the CLIP loss, the ``melting'' patterns visible in \textit{The Scream} are not synthesized as depicted in Figure~\ref{fig:ablations}.
On the other hand, the \textbf{NNFM loss} focuses on high-frequency patterns. 
For instance, the brush strokes are not as visible when switching off the NNFM loss, as depicted in Figure~\ref{fig:ablations}.
The pre-trained networks used for the style losses were not originally trained for this task but they were trained for classification. 
As such, using them for this purpose may cause them to produce noise-like artifacts (see Figure~\ref{fig:ablations}, no total variation loss). 
By using the \textbf{total variation loss}, we can remove these artifacts.
If an overly smooth appearance is desired, this can be achieved by using an even higher weight for this loss, which needs to be tuned for this specific purpose (see Figure~\ref{fig:ablations}, high total variation loss).

\section{Limitations}

By utilizing Gaussian Splatting as the underlying data representation, we also inherit certain visual artifacts that are either caused by their construction process or are fundamental to this representation.
Namely, when reconstructing a real-world scene, some ``stray'' Gaussians may appear as floaters, potentially inhibiting the stylization process.
\rev{This could be remedied by using a modified version of the original Gaussian Splatting approach~\cite{kerbl3Dgaussians}, which is currently an active area of research~\cite{chen2024gssurvey, fei2024gssurvey, wu2024gssurvey}.} 
Furthermore, splatting can lead to compositing artifacts.
Changing the viewpoint slightly may cause a Gaussian to pop in front of another one, given that during rendering the Gaussians are sorted based on the distance from their mean to the camera.
Moreover, the stylization process is unguided, which means that an artist \rev{lacks full control over the placement of style features.} Also, there is no straightforward way to ensure that patterns such as brush strokes point in the desired direction.
\rev{To the best of our knowledge, there are no reliable ways of selecting style patterns to reconstruct the content of a given image.
This seems to be an inherent limitation of style transfer methods based on transferring the distribution of high-level features across images.}
This could be remedied in future work by considering more advanced style transfer models that can be conditioned to ensure this degree of control.

Furthermore, we rely on pre-trained neural networks to extract features based on which we transfer styles.
However, there is no guarantee that the networks can truthfully capture the style features within their latent variables, which could lead to unconvincing results.
According to our results shown in Figures~\ref{fig:results_forwardOurs} and~\ref{fig:results_360Ours}, we are able to capture and synthesize the style of varied style images.
Yet, with a more robust architecture, we might be able to do so more closely to the original style.
Lastly, our approach relies on the stylization of 2D projected areas obtained with a differentiable renderer from a finite set of viewpoints.
Depending on the distribution of those viewpoints, certain areas may take longer to converge or might not be fully optimized.
\rev{Conducting an analysis to identify infrequently viewed areas and strategically placing new cameras in those locations could lead to faster convergence and improved results.}
\section{Conclusions}

We introduced a novel algorithm for stylizing a 3D scene represented by a set of Gaussians to match the style of a given image, $\mathscr{G}$-Style.
By optimizing the geometry of the scene based on the needs of the stylization process and by using a dual loss that captures high and low-frequency style patterns, we generate stylized scenes with higher quality than existing methods in a matter of minutes per style image.
In the future, we would like to address the aforementioned limitations, inherited from the Gaussian Splatting representation and the employed neural network architectures, aiming to further improve the quality of the stylized scenes.
\rev{Finally, we intend to perform additional editing experiments, such as blending multiple styles into one content image or conducting localized or semantic style transfer on distinct regions of the content image, to understand further the strengths and limitations of our proposed approach.}

\vspace{15pt}
\noindent\textbf{Acknowledgment} 
\rev{The authors would like to thank Dr. Bernhard Kerbl for his input at the early stages of this project. }

\bibliographystyle{eg-alpha-doi} 
\bibliography{egbibsample.bib}       

\newcommand{\etalchar}[1]{$^{#1}$}
\begin{thebibliography}{\uppercase{MWWL22}}

\bibitem[AHS{\etalchar{*}}21]{an2021artflowff}
\textsc{An J., Huang S., Song Y., Dou D., Liu W., Luo J.}:
\newblock Artflow: Unbiased image style transfer via reversible neural flows.
\newblock In \emph{Proceedings of the IEEE/CVF Conference on Computer Vision and Pattern Recognition} (2021).

\bibitem[CHH24]{chung2024style}
\textsc{Chung J., Hyun S., Heo J.-P.}:
\newblock Style injection in diffusion: A training-free approach for adapting large-scale diffusion models for style transfer.
\newblock In \emph{Proceedings of the IEEE/CVF International Conference on Computer Vision} (2024).

\bibitem[CS16]{chen2016styletransferfastnn}
\textsc{Chen T.~Q., Schmidt M.}:
\newblock Fast patch-based style transfer of arbitrary style.
\newblock In \emph{arXiv preprint arXiv:1612.04337} (2016).

\bibitem[CW10]{chen2010high}
\textsc{Chen J., Wang B.}:
\newblock High quality solid texture synthesis using position and index histogram matching.
\newblock \emph{The Visual Computer} (2010).

\bibitem[CW24]{chen2024gssurvey}
\textsc{Chen G., Wang W.}:
\newblock A survey on {3D} gaussian splatting.
\newblock \emph{arXiv preprint arXiv:2401.03890} (2024).

\bibitem[CWNN20]{cao2020styletransferpointclouds}
\textsc{Cao X., Wang W., Nagao K., Nakamura R.}:
\newblock {PSNet: A Style Transfer Network for Point Cloud Stylization on Geometry and Color}.
\newblock In \emph{2020 IEEE Winter Conference on Applications of Computer Vision (WACV)} (2020).

\bibitem[CXG{\etalchar{*}}22]{Chen2022ECCV}
\textsc{Chen A., Xu Z., Geiger A., Yu J., Su H.}:
\newblock Tensorf: Tensorial radiance fields.
\newblock In \emph{European Conference on Computer Vision (ECCV)} (2022).

\bibitem[FKYT{\etalchar{*}}22]{yu2022plenoxels}
\textsc{Fridovich-Keil S., Yu A., Tancik M., Chen Q., Recht B., Kanazawa A.}:
\newblock Plenoxels: Radiance fields without neural networks.
\newblock In \emph{Proceedings of the IEEE/CVF Conference on Computer Vision and Pattern Recognition} (2022).

\bibitem[FXZ{\etalchar{*}}24]{fei2024gssurvey}
\textsc{Fei B., Xu J., Zhang R., Zhou Q., Yang W., He Y.}:
\newblock {3D} gaussian splatting as new era: A survey.
\newblock \emph{IEEE Transactions on Visualization and Computer Graphics} (2024).

\bibitem[GCLY18]{gu2018arbitrarystyleloss}
\textsc{Gu S., Chen C., Liao J., Yuan L.}:
\newblock Arbitrary style transfer with deep feature reshuffle.
\newblock In \emph{Proceedings of the IEEE conference on computer vision and pattern recognition} (2018).

\bibitem[GEB15a]{gatys2015styletransfer}
\textsc{Gatys L.~A., Ecker A.~S., Bethge M.}:
\newblock A neural algorithm of artistic style.
\newblock \emph{arXiv preprint arXiv:1508.06576} (2015).

\bibitem[GEB15b]{gatys2015texturesynthesis}
\textsc{Gatys L.~A., Ecker A.~S., Bethge M.}:
\newblock Texture synthesis using convolutional neural networks.
\newblock \emph{Advances in neural information processing systems 28} (2015).

\bibitem[GRGH19]{gutierrez2019volumetexture}
\textsc{Gutierrez J., Rabin J., Galerne B., Hurtut T.}:
\newblock {On Demand Solid Texture Synthesis Using Deep {3D} Networks}.
\newblock \emph{Computer Graphics Forum} (2019).

\bibitem[HB17]{huang2017arbitrarytransferff}
\textsc{Huang X., Belongie S.}:
\newblock Arbitrary style transfer in real-time with adaptive instance normalization.
\newblock In \emph{Proceedings of the IEEE international conference on computer vision} (2017).

\bibitem[HHY{\etalchar{*}}22]{Huang_2022_CVPR}
\textsc{Huang Y.-H., He Y., Yuan Y.-J., Lai Y.-K., Gao L.}:
\newblock Stylizednerf: Consistent {3D} scene stylization as stylized nerf via {2D}-{3D} mutual learning.
\newblock In \emph{Proceedings of the IEEE/CVF Conference on Computer Vision and Pattern Recognition (CVPR)} (2022).

\bibitem[HJN22]{hollein2022styletransferrender}
\textsc{H{\"o}llein L., Johnson J., Nie{\ss}ner M.}:
\newblock {Stylemesh: Style Transfer for Indoor {3D} Scene Reconstructions}.
\newblock In \emph{Proceedings of the IEEE/CVF Conference on Computer Vision and Pattern Recognition (CVPR)} (2022).

\bibitem[HMR20]{henzler2020neuraltexture}
\textsc{Henzler P., Mitra N.~J., Ritschel T.}:
\newblock {Learning a Neural {3D} Texture Space from {2D} Exemplars}.
\newblock In \emph{The IEEE Conference on Computer Vision and Pattern Recognition (CVPR)} (2020).

\bibitem[HPP{\etalchar{*}}18]{hedman2018deep}
\textsc{Hedman P., Philip J., Price T., Frahm J.-M., Drettakis G., Brostow G.}:
\newblock Deep blending for free-viewpoint image-based rendering.
\newblock \emph{ACM Transactions on Graphics (ToG)} (2018).

\bibitem[HTS{\etalchar{*}}21]{Huang_2021_ICCV}
\textsc{Huang H.-P., Tseng H.-Y., Saini S., Singh M., Yang M.-H.}:
\newblock Learning to stylize novel views.
\newblock In \emph{Proceedings of the IEEE/CVF International Conference on Computer Vision (ICCV)} (2021).

\bibitem[JBV17]{jetchev2017texturesynthesisgan}
\textsc{Jetchev N., Bergmann U., Vollgraf R.}:
\newblock Texture synthesis with spatial generative adversarial networks.
\newblock \emph{arXiv preprint arXiv:1611.08207} (2017).

\bibitem[{Ker}23]{3dgs_code}
\textsc{{Kerbl, Bernhard and Kopanas, Georgios and Leimk{\"u}hler, Thomas and Drettakis, George}}:
\newblock {{3D} Gaussian Splatting for Real-Time Radiance Field Rendering --- {Implementation}}.
\newblock \url{https://github.com/graphdeco-inria/gaussian-splatting}, 2023.

\bibitem[KFCO{\etalchar{*}}07]{kopf2007solid}
\textsc{Kopf J., Fu C.-W., Cohen-Or D., Deussen O., Lischinski D., Wong T.-T.}:
\newblock {Solid Texture Synthesis from {2D} Exemplars}.
\newblock In \emph{ACM SIGGRAPH 2007}. ACM, 2007.

\bibitem[KHR24]{kovacs2024surface}
\textsc{Kov{\'a}cs {\'A}.~S., Hermosilla P., Raidou R.~G.}:
\newblock Surface-aware mesh texture synthesis with pre-trained {2D} cnns.
\newblock In \emph{Computer Graphics Forum (Eurographics)} (2024).

\bibitem[KKLD23]{kerbl3Dgaussians}
\textsc{Kerbl B., Kopanas G., Leimk{\"u}hler T., Drettakis G.}:
\newblock {3D} gaussian splatting for real-time radiance field rendering.
\newblock \emph{ACM Transactions on Graphics} (2023).

\bibitem[KPZK17]{Knapitsch2017}
\textsc{Knapitsch A., Park J., Zhou Q.-Y., Koltun V.}:
\newblock Tanks and temples: Benchmarking large-scale scene reconstruction.
\newblock \emph{ACM Transactions on Graphics} (2017).

\bibitem[KSS19]{kolkin2019styleoptimaltransportlossnn}
\textsc{Kolkin N., Salavon J., Shakhnarovich G.}:
\newblock Style transfer by relaxed optimal transport and self-similarity.
\newblock In \emph{Proceedings of the IEEE/CVF conference on computer vision and pattern recognition} (2019).

\bibitem[LW16]{li2016combiningmarkovnn}
\textsc{Li C., Wand M.}:
\newblock Combining markov random fields and convolutional neural networks for image synthesis.
\newblock \emph{IEEE Conference on Computer Vision and Pattern Recognition (CVPR)} (2016).

\bibitem[LYY{\etalchar{*}}17]{liao2017imageanalogylossnn}
\textsc{Liao J., Yao Y., Yuan L., Hua G., Kang S.~B.}:
\newblock Visual attribute transfer through deep image analogy.
\newblock \emph{ACM Transactions on Graphics} (2017).

\bibitem[LZC{\etalchar{*}}23a]{liu2023stylerf}
\textsc{Liu K., Zhan F., Chen Y., Zhang J., Yu Y., Saddik A.~E., Lu S., Xing E.}:
\newblock Stylerf: Zero-shot {3D} style transfer of neural radiance fields.
\newblock \emph{Proc. IEEE Conf. on Computer Vision and Pattern Recognition (CVPR)} (2023).

\bibitem[LZC{\etalchar{*}}23b]{stylerf_code}
\textsc{Liu K., Zhan F., Chen Y., Zhang J., Yu Y., Saddik A.~E., Lu S., Xing E.}:
\newblock Stylerf: Zero-shot {3D} style transfer of neural radiance fields --- {Implementation}.
\newblock \url{https://github.com/Kunhao-Liu/StyleRF}, 2023.

\bibitem[LZL{\etalchar{*}}23]{liu2023instant}
\textsc{Liu R., Zhao E., Liu Z., Feng A., Easley S.~J.}:
\newblock Instant photorealistic style transfer: A lightweight and adaptive approach.
\newblock \emph{arXiv preprint arXiv:2309.10011} (2023).

\bibitem[LZX{\etalchar{*}}24]{liu2023stylegaussian}
\textsc{Liu K., Zhan F., Xu M., Theobalt C., Shao L., Lu S.}:
\newblock {StyleGaussian: Instant {3D} Style Transfer with Gaussian Splatting}.
\newblock \emph{arXiv preprint arXiv:2403.07807} (2024).

\bibitem[Max95]{max1995optical}
\textsc{Max N.}:
\newblock Optical models for direct volume rendering.
\newblock \emph{IEEE Transactions on Visualization and Computer Graphics} (1995).

\bibitem[MPSO18]{mordvintsev2018styletransferrender}
\textsc{Mordvintsev A., Pezzotti N., Schubert L., Olah C.}:
\newblock Differentiable image parameterizations.
\newblock \emph{Distill 3}, 7 (2018), e12.

\bibitem[MSOC{\etalchar{*}}19]{mildenhall2019llff}
\textsc{Mildenhall B., Srinivasan P.~P., Ortiz-Cayon R., Kalantari N.~K., Ramamoorthi R., Ng R., Kar A.}:
\newblock Local light field fusion: Practical view synthesis with prescriptive sampling guidelines.
\newblock \emph{ACM Transactions on Graphics (TOG)} (2019).

\bibitem[MST{\etalchar{*}}20]{mildenhall2020nerf}
\textsc{Mildenhall B., Srinivasan P.~P., Tancik M., Barron J.~T., Ramamoorthi R., Ng R.}:
\newblock {NeRF: Representing Scenes as Neural Radiance Fields for View Synthesis}.
\newblock In \emph{European Conference on Computer Vision (ECCV)} (2020).

\bibitem[MWWL22]{Mu_2022_CVPR}
\textsc{Mu F., Wang J., Wu Y., Li Y.}:
\newblock {3D} photo stylization: Learning to generate stylized novel views from a single image.
\newblock In \emph{Proceedings of the IEEE/CVF Conference on Computer Vision and Pattern Recognition (CVPR)} (2022).

\bibitem[NPLX22]{nguyen2022snerf}
\textsc{Nguyen-Phuoc T., Liu F., Xiao L.}:
\newblock {Snerf: stylized neural implicit representations for {3D} scenes}.
\newblock \emph{SIGGRAPH} (2022).

\bibitem[RKH{\etalchar{*}}21]{radford2021learning}
\textsc{Radford A., Kim J.~W., Hallacy C., Ramesh A., Goh G., Agarwal S., Sastry G., Askell A., Mishkin P., Clark J., et~al.}:
\newblock Learning transferable visual models from natural language supervision.
\newblock In \emph{International conference on machine learning} (2021).

\bibitem[SGC{\etalchar{*}}24]{saroha2024gaussian}
\textsc{Saroha A., Gladkova M., Curreli C., Yenamandra T., Cremers D.}:
\newblock Gaussian splatting in style.
\newblock \emph{arXiv preprint arXiv:2403.08498} (2024).

\bibitem[SZ14]{simonyan2014vgg16}
\textsc{Simonyan K., Zisserman A.}:
\newblock Very deep convolutional networks for large-scale image recognition.
\newblock \emph{International Conference on Learning Representations (ICLR)} (2014).

\bibitem[Ull79]{ullman1979interpretation}
\textsc{Ullman S.}:
\newblock The interpretation of structure from motion.
\newblock \emph{Proceedings of the Royal Society of London. Series B. Biological Sciences} (1979).

\bibitem[WYZ{\etalchar{*}}24]{wu2024gssurvey}
\textsc{Wu T., Yuan Y.-J., Zhang L.-X., Yang J., Cao Y.-P., Yan L.-Q., Gao L.}:
\newblock Recent advances in {3D} gaussian splatting.
\newblock \emph{Computational Visual Media} (2024), 1--30.

\bibitem[WZX23]{wang2023stylediffusion}
\textsc{Wang Z., Zhao L., Xing W.}:
\newblock Stylediffusion: Controllable disentangled style transfer via diffusion models.
\newblock In \emph{Proceedings of the IEEE/CVF International Conference on Computer Vision} (2023).

\bibitem[XCX{\etalchar{*}}24]{xu2024styledyrf}
\textsc{Xu H., Chen W., Xiao F., Sun B., Kang W.}:
\newblock {StyleDyRF: Zero-shot 4D Style Transfer for Dynamic Neural Radiance Fields}.
\newblock \emph{arXiv preprint arXiv:2403.08310} (2024).

\bibitem[ZCY{\etalchar{*}}24]{zhang2024stylizedgs}
\textsc{Zhang D., Chen Z., Yuan Y.-J., Zhang F.-L., He Z., Shan S., Gao L.}:
\newblock {StylizedGS: Controllable Stylization for {3D} Gaussian Splatting}.
\newblock \emph{arXiv preprint arXiv:2404.05220} (2024).

\bibitem[ZFLS24]{zhang2024coarf}
\textsc{Zhang D., Fernandez-Labrador C., Schroers C.}:
\newblock {CoARF: Controllable {3D} Artistic Style Transfer for Radiance Fields}.
\newblock \emph{Internation Conference on {3D} Vision ({3D}V)} (2024).

\bibitem[ZGW{\etalchar{*}}22]{zhao2022stsgan}
\textsc{Zhao X., Guo J., Wang L., Li F., Zheng J., Yang B.}:
\newblock {STS-GAN: Can We Synthesize Solid Texture with High Fidelity from Arbitrary Exemplars?}
\newblock \emph{Proceedings of the Thirty-Second International Joint Conference on Artificial Intelligence} (2022).

\bibitem[ZHT{\etalchar{*}}23]{zhang2023diffusionstyletransfer}
\textsc{Zhang Y., Huang N., Tang F., Huang H., Ma C., Dong W., Xu C.}:
\newblock Inversion-based style transfer with diffusion models.
\newblock In \emph{Proceedings of the IEEE/CVF Conference on Computer Vision and Pattern Recognition (CVPR)} (2023).

\bibitem[ZKB{\etalchar{*}}22a]{zhang2022arf}
\textsc{Zhang K., Kolkin N., Bi S., Luan F., Xu Z., Shechtman E., Snavely N.}:
\newblock {ARF: Artistic radiance fields}.
\newblock In \emph{European Conference on Computer Vision} (2022).

\bibitem[ZKB{\etalchar{*}}22b]{arf_code}
\textsc{Zhang K., Kolkin N., Bi S., Luan F., Xu Z., Shechtman E., Snavely N.}:
\newblock {ARF: Artistic radiance fields --- {Implementation}}.
\newblock \url{https://github.com/Kai-46/ARF-svox2}, 2022.

\end{thebibliography}

\end{document}